\documentclass[letterpaper, 10 pt, conference]{ieeeconf}  

\IEEEoverridecommandlockouts                              

\usepackage{subcaption}
\usepackage{cite}
\usepackage{amsmath,amssymb,amsfonts}
\usepackage{algorithmic}
\usepackage{graphicx}
\usepackage{textcomp}
\usepackage{url}
\usepackage{xcolor}
\usepackage{siunitx} \usepackage{flushend}
\usepackage{tikz}
\usetikzlibrary{positioning, arrows.meta,automata}
\usetikzlibrary{calc}

\newtheorem{problem}{Problem}

\def \x{\mathbf{x}}
\def \Q{\mathcal{Q}}
\def \W{\mathcal{W}}

\def \T{\mathcal{T}}
\def \V{\mathcal{V}}
\def \E{\mathcal{E}}
\def \Q{\mathcal{Q}}

\title{\LARGE \bf NNgTL: Neural Network Guided Optimal Temporal Logic\\ Task Planning for Mobile Robots
}

\author{Ruijia Liu, Shaoyuan Li and Xiang Yin
\thanks{This work was supported by the National Natural Science Foundation of China (62061136004, 62173226, 61803259).)
}
\thanks{R. Liu, S. Li and  X. Yin are with the Department of Automation, Shanghai Jiao Tong University, Shanghai 200240, China, and also with the Key Laboratory of System Control and Information Processing, the Ministry of Education of China, Shanghai 200240, China. {\tt  E-mail: \{liuruijia,syli,yinxiang\}@sjtu.edu.cn}} 
}

\begin{document}

\maketitle
\thispagestyle{empty}
\pagestyle{empty}
\setlength{\abovecaptionskip}{3pt}
\setlength{\belowcaptionskip}{0pt}
\setlength{\textfloatsep}{6pt}

\begin{abstract}
In this work, we investigate task planning for mobile robots  under linear temporal logic (LTL) specifications. This problem is particularly challenging when robots navigate in continuous workspaces due to the high computational complexity involved. Sampling-based methods have emerged as a promising avenue for addressing this challenge   by incrementally constructing random trees, thereby sidestepping the need to explicitly explore the entire state-space. However, the 
performance of this sampling-based approach hinges crucially on the chosen sampling strategy, and a well-informed heuristic can notably enhance sample efficiency. 
In this work, we propose a novel \emph{neural-network guided} (NN-guided) sampling  strategy tailored for LTL planning. Specifically, we employ a multi-modal neural network capable of extracting features concurrently from both the workspace and the B\"{u}chi automaton. This neural network generates predictions that serve as guidance for random tree construction, directing the sampling process toward more optimal directions.
Through numerical experiments, we compare our approach with existing methods and demonstrate its superior efficiency, requiring less than 15\% of the time of the existing methods to find a feasible solution. 
\end{abstract}


\section{Introduction}
With the ongoing development and widespread deployment of mobile robots, there has been an increasing focus on path planning for high-level tasks. Among the formal languages used for specifying such complex tasks, Linear Temporal Logic (LTL) stands out as a widely adopted choice. LTL provides a structured means for the users to articulate complex  requirements, such as navigating a robot to a target region without collision with obstacles or ensuring specific locations are visited infinitely often \cite{guo2018probabilistic,cai2022learning,luo2022temporal,yu2022security}. In recent years, the field of robot path planning for LTL tasks has witnessed extensive investigation,  spurred by its broad  applications. These applications encompass a wide range of scenarios, such as environmental surveillance \cite{smith2011optimal}, search and rescue missions \cite{plaku2016motion}, and   intelligent warehousing systems \cite{yang2019formal,scher2020warehouse}.

In the context of LTL path planning,  one of the most foundational methods is the automata-theoretical approach  based on finite abstractions  \cite{smith2011optimal,guo2015multi,kloetzer2009automatic,shi2022path,ulusoy2014optimal}. 
This approach involves the creation of a discrete abstraction of the workspace, which effectively captures the robot's mobility constraints. Subsequently, the discrete abstraction is synchronized with an automaton representation of the LTL task. This synchronization enables the formulation of the planning problem as a graph-search problem within the product space.
However, this graph-search approach, although conceptually powerful, faces a significant computational challenge. In particular, as the system's dimensionality increases, the state-space of the finite abstraction grows exponentially, rendering the graph-search problem infeasible.

Sampling-based methods, such as rapid random trees (RRT), have emerged as a promising solution to tackle the computational challenges associated with path planning in continuous state-spaces  \cite{karaman2011sampling}. More recently,   sampling-based algorithms have been introduced to enhance the computational efficiency of solving LTL planning problems \cite{karaman2012sampling,vasile2013sampling,kantaros2018sampling,kantaros2020stylus,luo2021abstraction}.
For example,  in \cite{kantaros2018sampling}, a sampling-based algorithm, inspired by the RRT*, is proposed to find optimal paths that satisfy LTL tasks with a prefix-suffix structure. This  algorithm circumvents the necessity of explicitly exploring the entire state-space by incrementally constructing random trees over the product state-space. Building upon this advancement,  \cite{kantaros2020stylus} further introduces a biased sampling strategy that leverages automata information to significantly improve sample efficiency. Moreover, in  \cite{luo2021abstraction}, the authors extend the methods in \cite{kantaros2020stylus} to continuous workspaces without discretization. Specifically, they introduce an abstraction-free planning algorithm that directly samples within a continuous space and integrates these samples with automata states.

In sampling-based LTL planning, one of the key factor is how each new state is sampled.  While biased sampling strategies offer an enhanced approach compared to uniform sampling, they primarily rely on distance information within the B\"{u}chi automaton. This approach neglects valuable insights from the workspace, such as the physical feasibility of task progression. However, integrating continuous workspace information with the B\"{u}chi automaton  is very challenging  due to the inherently heterogeneous structures of these components. Recently, there has been a notable 
progress in leveraging the power of neural networks for expediting solutions to intricate problems. For instance, within the domain of path planning, neural networks have been employed to predict the probability distribution of optimal paths  \cite{ichter2018learning,wang2020neural,ma2021conditional,qureshi2020motion}. Moreover, neural networks have been applied, in an end-to-end fashion, for generating solutions to temporal logic tasks  \cite{liu2021recurrent,hashimoto2022stl2vec,cai2021modular,aloor2023follow}. 

In this paper, we introduce a novel neural-network guided (NN-guided) sampling strategy designed specifically for LTL planning. Our approach builds upon the basic architecture of the sampling-based algorithm   for continuous workspaces, as established in \cite{luo2021abstraction}. However, instead of relying solely on automata information, we incorporate a multi-modal neural network capable of jointly extracting features from both the workspace and the B\"{u}chi automaton. This neural network offers predictive capabilities to steer the sampling process in directions that are more likely to yield optimal solutions, ones that are not only task-progressive but also feasible within the workspace. 
To demonstrate the efficiency of our approach, 
we compare the proposed NN-guided sampling strategy with existing sampling strategies on a set of randomly generated problem instances. 
The statistical results underscore the effectiveness of our new strategy, as it requires less than 15\% of the time needed by the existing strategies to find a feasible solution.

\section{Problem Formulation}

\subsection{System Model} 
We consider the scenario, where a single robot navigating in a two-dimensional continuous workspace  represented by a compact subset $\mathcal{W}\subseteq \mathbb{R}^2$. 
The workspace is assumed to be partitioned as $\mathcal{W}=\mathcal{W}_\text{free}\dot{\cup}\mathcal{O}$, 
where $\mathcal{O}$ is an open subset representing obstacle regions and $\mathcal{W}_\text{free}$ is the free region in which the robot can move.  
The free workspace is further partitioned as  $m$ labeled regions of interest and a non-labeled region
$\mathcal{W}_\text{free}=\mathcal{R}_1\dot{\cup} \cdots\dot{\cup} \mathcal{R}_m\dot{\cup} \mathcal{R}_{non}$. 
 
The mobility of the robot can be captured by  a weighted transition system 
\[
TS = (\mathcal{W}, \mathbf{x}_0, \rightarrow_T, \mathcal{A P}, L, C),
\]
where 
$\mathcal{W}$ is set of infinite positions of the robot; 
$\mathbf{x}_0$ is its initial position; 
$\rightarrow_T \subseteq \mathcal{W} \times \mathcal{W}$ is the transition relation such that: 
for any $\x,\x'\in \mathcal{W}$, we have $(\x,\x')\in \to_T $ if along the straight line between $\x$ and $\x'$ 
(i) does not intersect with $\mathcal{O}$; and  
(ii)  crosses any  boundary of labeled regions at most once \cite{luo2021abstraction};  
$\mathcal{A P}=\{\pi^{1},\dots, \pi^{m}\}$ is the set of atomic propositions
with labeling function $L: \mathcal{W} \to \mathcal{A P}$ such that $L(\x)=\pi^{i}$ iff $\x\in \mathcal{R}_i$; 
$C: \mathcal{W} \times \mathcal{W} \to \mathbb{R}^{+}$ is the weight function  calculating  the Euclidean distance, i.e., $C(\mathbf{x}_1, \mathbf{x}_2) = \left\|\mathbf{x}_1 - \mathbf{x}_2\right\|_2$.

An infinite \emph{path} is a sequence $\tau \!=\! \tau(1) \tau(2) \cdots\! \in\! \mathcal{W}^\omega$ such that 
$\tau(1) = \x_0$ and  $(\tau(k),\tau(k+1))\in \to_T, k=1,2,\dots$.  
The \emph{trace} of infinite path $\tau$ is  $\operatorname{trace}(\tau) = L(\tau(1)) L(\tau(2)) \cdots \in (\mathcal{AP})^\omega$. 
We say a path $\tau$ is finite if it is a finite sequence of points, and we denote by $|\tau|$ its \emph{length}. 
For  a finite path $\tau$, its cost $J(\tau)$ is defined as the the cumulative distances between consecutive states, 
i.e.,  $J(\tau) = \sum_{k=1}^{|\tau|-1} C(\tau(k), \tau(k+1))$.

\subsection{Temporal Logic Tasks}
The formal specification of robot is captured by an linear temporal logic formula without next (LTL$_{-\bigcirc}$), which is widely used in robot task planning in continuous workspace \cite{kloetzer2008fully}. 
The syntax of LTL$_{-\bigcirc}$ is given as follows: 
\[
\phi::=\text{true} \mid  \pi\in \mathcal{A P}\mid   \neg \phi  \mid \phi_1 \wedge \phi_2  \mid   \phi_1  \mathcal{U}   \phi_2, 
\]
where $\neg$ and $\wedge$ are Boolean operators ``negation" and ``conjunction", respectively; 
$\mathcal{U}$ is the temporal operator ``until", which can further induce temporal operators ``eventually" $\Diamond$ and ``always" $\square$. 

An LTL formula $\phi$ is evaluated over an infinite word $\sigma=\sigma(1)\sigma(2)\cdots\in \mathcal{AP}^\omega$. 
We denote by $\sigma\models \phi$ that word $\sigma$ satisfies LTL formula $\phi$; the reader is referred to \cite{baier2008principles} for the formal semantics. 
We denote by $\operatorname{Words}(\phi)$   the set of all words   satisfying  formula $\phi$. 
It is well-known that, for any LTL formula $\phi$, $\operatorname{Words}(\phi)$ can be accepted by a non-deterministic B\"{u}chi automaton (NBA) \cite{baier2008principles}.  
Formally, an NBA is a tuple $B=(\mathcal{Q}_B, \mathcal{Q}_B^0,\Sigma, \rightarrow_B, \mathcal{Q}_B^F)$, 
where 
$\mathcal{Q}_B$ is the set of states, 
$\mathcal{Q}_B^0$ is the set of initial states, 
$\Sigma =\mathcal{AP}$ is the alphabet, 
$\rightarrow_B \subseteq \mathcal{Q}_B \times \Sigma \times \mathcal{Q}_B$ is the transition relation, 
and $\mathcal{Q}_B^F$ is the set of accepting states. 
For simplicity, we assume that the initial-state is unique, i.e.,  $\mathcal{Q}_B^0=\{q_B^0\}$. 

An infinite run $\rho_B$ of $B$ over an infinite word $\sigma=\pi_0 \pi_1   \cdots \in\left( {\mathcal{A P}}\right)^\omega$  is a sequence $\rho_B=q_B^0 q_B^1 q_B^2 \cdots$ such that  $q_B^0 \in \mathcal{Q}_B^0,\left(q_B^i, \pi_i, q_B^{i+1}\right) \in \rightarrow_B, \forall i=0,1,\dots$. We say infinite run $\rho_B$  is accepting if it contains accepting states infinite number of times; 
we say a word is accepting if it induces an accepting run, and we denote by $\mathcal{L}_B$ the set of all accepting words of $B$. 
Hereafter, $B$ is referred to as the NBA that accepts $\phi$, i.e., $\mathcal{L}_B = \operatorname{Words}(\phi)$. 

\subsection{LTL Task Planning Problem}
To fulfill an  $\mathrm{LTL}_{-\bigcirc}$ formula $\phi$, the robot needs to execute an infinite path. 
For the purpose of planning, it suffices to consider paths in the   ``prefix-suffix'' structure 
$\tau = \tau^{\mathrm{pre}}[\tau^{\mathrm{suf}}]^\omega$, 
where the prefix part $\tau^{\mathrm{pre}}$ repeats only once and the suffix part $\tau^{\mathrm{suf}}$ repeats indefinitely \cite{guo2015multi}. The cost of   prefix-suffix path $\tau$ is  defined by 
$J(\tau)=\lambda J(\tau^{\mathrm{pre}})+(1-\lambda)J(\tau^{\mathrm{suf}})$, where $\lambda \in [0,1]$ is a user-defined weight coefficient.

Then our objective is to find a plan for the robot in the prefix-suffix structure such that  
a given LTL formula is fulfilled with minimum cost.  
 
\begin{problem}\label{problem1}
Given $\mathrm{LTL}_{-\bigcirc}$ formula $\phi$, determine a prefix-suffix path $\tau$ in   transition system $TS$ 
such that (i) $\operatorname{trace}(\tau)\models \phi$; and 
(ii) for any prefix-suffix path $\tau'$ such that $\operatorname{trace}(\tau')\models \phi$, 
we have $J(\tau)\leq J(\tau')$. 
\end{problem}

\section{Sampling-Based Task Planning}\label{section_heuristicsample}
To  solve  Problem \ref{problem1}, a typical approach is to perform graph-search  on the \emph{product} of $TS$ and $B$ \cite{smith2011optimal}. 
However, when the state space $\mathcal{W}$ is continuous, building  the entire product space is infeasible even by discretizations.  
Therefore, in \cite{luo2021abstraction}, the authors proposed a (continuous space) sampling-based algorithm, called TL-RRT*, that incrementally builds trees on-the-fly without construct  the entire product space a priori. 
Our work is build upon the sampling-based approach; therefore, we review it briefly in this section. 
The readers are referred to \cite{luo2021abstraction} for more details on this method. 

\subsection{Main Sampling-Based Algorithm} 
The TL-RRT* algorithm is essentially a random search over the product space $\Q_P:=\mathcal{W}_{\text{free}}\times \Q_B$. It consists of two parts, prefix part and suffix part, and for each part, a similar random tree construction is employed in order to search the optimal path. We briefly sketch the procedures. 
 
\subsubsection{Prefix Search}
Starting from the initial state $q_0=(\x_0, q_B^0)$, one first builds a prefix tree $\T=( \mathcal{V}_{\mathcal{T}},\mathcal{E}_{\mathcal{T}},\operatorname{Cost} )$ by incrementally adding vertices, where $\mathcal{V}_{\mathcal{T}}$ and $\E_\T$ are the sets of vertices and edges in $\T$, respectively.  
The tree contains some  \emph{goal states}  defined by
    $\mathcal{Q}_{\text {goal}}:=  \mathcal{W}_{\text{free}}\times  \mathcal{Q}_B^F$.  
By projecting paths in tree onto $\mathcal{W}$,  
the path from the initial state $q_0$ to each goal state 
    $q_{\text{goal}}\in\V_\T\cap \mathcal{Q}_{\text {goal}}$ forms a prefix path $\tau^{\mathrm{pre}}$, whose cost is stored by function $\text {Cost}: \mathcal{V}_{\mathcal{T}} \rightarrow \mathbb{R}^{+}$. 

\subsubsection{Suffix Search}
The suffix part is very similar to the prefix part, 
the main difference is that one needs to construct a set of random trees rooted from  goal states $\mathcal{P}:=\T\cap \mathcal{Q}_{\text {goal}}$ in the prefix tree
such that they can return back to the goal states. 
This gives us a suffix  plan  $\tau^{\mathrm{suf}}$ from  each goal state,  and the prefix-suffix plan with minimum cost among all goal states is the final optimal plan.

\subsubsection{Constructing Random Trees}
In the above two parts, the key is construction of random tree  $\T=( \mathcal{V}_{\mathcal{T}},\mathcal{E}_{\mathcal{T}},\operatorname{Cost} )$ on-the-fly. The main construction steps are  as follows. 
\begin{enumerate}
    \item 
    Sample a state $ \mathbf{x}^{\text {rand}}\in  \mathcal{W}_{\text {free }}$ ``randomly"; 
    \item
    Determine a  new state $ \mathbf{x}^{\text {new}} $ to be added to the tree based on some distance criteria 
    on $ \mathbf{x}^{\text {rand}} $ and $\T$;  
    \item 
    Determine set  $ \Q^{\text {near}} \subseteq \V_\T$  based on some distance criteria
    from which the tree will be extended to $\x^{\text {new}}$; 
    \item 
    For each state 
    $ q_{P}^{\text {near }}=(\mathbf{x}^{\text{near}},q_B^{\text{near}}) \in  \mathcal{Q}_{P}^{\text {near }} $, 
    we consider a potential edge from  
    $q_P$  to  $ q_{P}^{\text {new}}=\left(\mathbf{x}^{\text {new}}, q_B\right) $ 
    if 
    $ (\x^{\text{near}} , \x^{\text{new}})\in \rightarrow_T $ 
    and $(q_B^{\text {near }}, L (\x^{\text {near }}), q_B^{\text {new }}) \in \rightarrow_B $;   
    \item 
    Add $q_{P}^{\text {new}}$ to tree $\T$ from state $q_P\in \mathcal{Q}_{P}^{\text {near }}$ to minimize  $ \operatorname{Cost}(q_{P}^{\text {new }})\!=\!\operatorname{Cost}(q_{P}^{\text {near}})\!+\!C(q_{P}^{\text {near}},q_{P}^{\text {new}}) $; 
    \item 
    After adding $q_{P}^{\text {new}}$, the tree edges and costs are reconfigured so that each vertex is reached by a path with minimum cost from the root.
\end{enumerate} 

\subsection{Sampling Strategies} 
In the construction of random trees, it remains to specify how state $\x^{\text{rand}}$ is sampled ``randomly".  
In \cite{luo2021abstraction}, the authors proposed two different sampling strategies for $\x^{\text{rand}}$: 
\begin{itemize}
    \item 
    \emph{Uniform Sampling: } 
    $\x^{\text{rand}}$ is selected according to a uniform distribution   on $\W_\text{free}$ 
    (in fact,  any distribution  as long as all states has non-zero probability);
    \item 
    \emph{Biased Sampling: }  
    $\x^{\text{rand}}$ is selected according to a biased distribution on $\W_\text{free}$
    such that one has more chance to move towards an accepting state in $B$. 
\end{itemize}
Since our NN-guided approach is related to the biased sampling, we briefly review this method. 
First, one needs to pre-process the NBA $B$ so that infeasible transitions are removed. 
Then based on the simplified NBA, 
one defines a distance function $\rho: \mathcal{Q}_B \times \mathcal{Q}_B \rightarrow \mathbb{N}$ to capture the length of the shortest path between each pair of states. 
Then the selection of $\x^{\text{rand}}$ is determined as follows:
\begin{enumerate}
    \item[S-1] 
    Select a \emph{feasible accepting state} $q_B^{F,\mathrm{feas}}\in\Q_B^F$ such that $\rho(q_B^0,q_B^{F,\mathrm{feas}})\neq \infty $ and $\rho(q_B^{F,\mathrm{feas}},q_B^{F,\mathrm{feas}})\neq \infty$;
    \item[S-2] 
    Define the set of vertices 
    $\mathcal{D}_\mathrm{min}\subseteq \mathcal{V}_{\mathcal{T}}$ that are closest to $q_B^{F,\mathrm{feas}}$  according the distance function $\rho$.  
    Then select a vertex 
    $q_{P}^{\mathrm {closest}}=\left(\mathbf{x}^{\mathrm {closest }}, q_{B}^{\mathrm {closest }}\right)\in \mathcal{V}_{\mathcal{T}}$ from the tree 
    according to a specific  distribution \cite{luo2021abstraction} such that states in $\mathcal{D}_\mathrm{min}$ has more chance to be selected;
    \item[S-3] 
    Select two successive states $q_{B}^{\text{succ,1}}, q_{B}^{\text{succ,2}}\in \Q_B$   such that $q_{B}^{\text{closest}} 
    \!\!\xrightarrow{L\left(\mathbf{x}^{\text {closest }}\right)}_B\! \! q_{B}^{\text{succ},1}\! \to_{B}\! q_{B}^{\text{succ},2}$, and
    $\rho(q_{B}^{\text{succ},2},q_B^{\mathrm{feas}})\!\leq\! \rho(q_{B}^{\text{succ},1},q_B^{F,\mathrm{feas}}) \!\leq\!  \rho(q_{B}^{\text{closest}},q_B^{F,\mathrm{feas}})$;
    \item[S-4] 
    Select a state $ \mathbf{x}^\mathcal{L}$ 
    such that  $ q_{B}^{\text{succ,1}}  \xrightarrow{L(\mathbf{x}^\mathcal{L})}_B q_{B}^{\text{succ, 2}} $ is feasible;  
    \item[S-5] 
    Compute the shortest path from 
    $\mathbf{x}^\mathcal{L} $ and $ \mathbf{x}^{\mathrm {closest}} $ within $ \mathcal{W}_{\text{free}} $. 
    Pick the second point in the shortest path, 
    denoted by $ \mathbf{x}^{\mathrm {target}} $
    as a heuristic direction for sampling;  
    \item[S-6] 
    Finally, 
    select $\x^{\text{rand}}$ according to a distribution  that has more chance to sample towards the direction of $\x^{\text{target}}$.
\end{enumerate}

\section{Neural Network Guided Sampling}
%
Although biased sampling provides a more efficient approach compared with the uniform sampling,  it solely relies on the distance information in B\"{u}chi automaton.  
For example, when  $q_B^{F,\mathrm{feas}}$ 
as well as 
intermediate states
$q_{B}^{\text{succ,1}}$  and $q_{B}^{\text{succ,2}}$
are selected, one only consider 
whether or not the task progress can be pushed forward in the NBA. 
However,  the information from the workspace, i.e., whether the task progress is feasible physically, is neglected.

To further improve the sample efficiency for the random tree construction,  one essentially needs to \emph{jointly} consider the information workspace and the NBA in order to have more chance to sample states that can move towards accepting states \emph{by paths that are feasible in the workspace}.  
However, obtaining such a good heuristic is very challenging since the workspace is continuous.  In this section, we present a new neural network guided (NN-guided) sampling strategy,  where a ``sample net" is used that effectively fuses the information of the workspace and the NBA.  
All implementation details as well as source codes are available at \url{https://github.com/LRJ-2000/NNgTL}

\begin{figure*}[htbp]
    \centering
    \includegraphics[width=17cm]{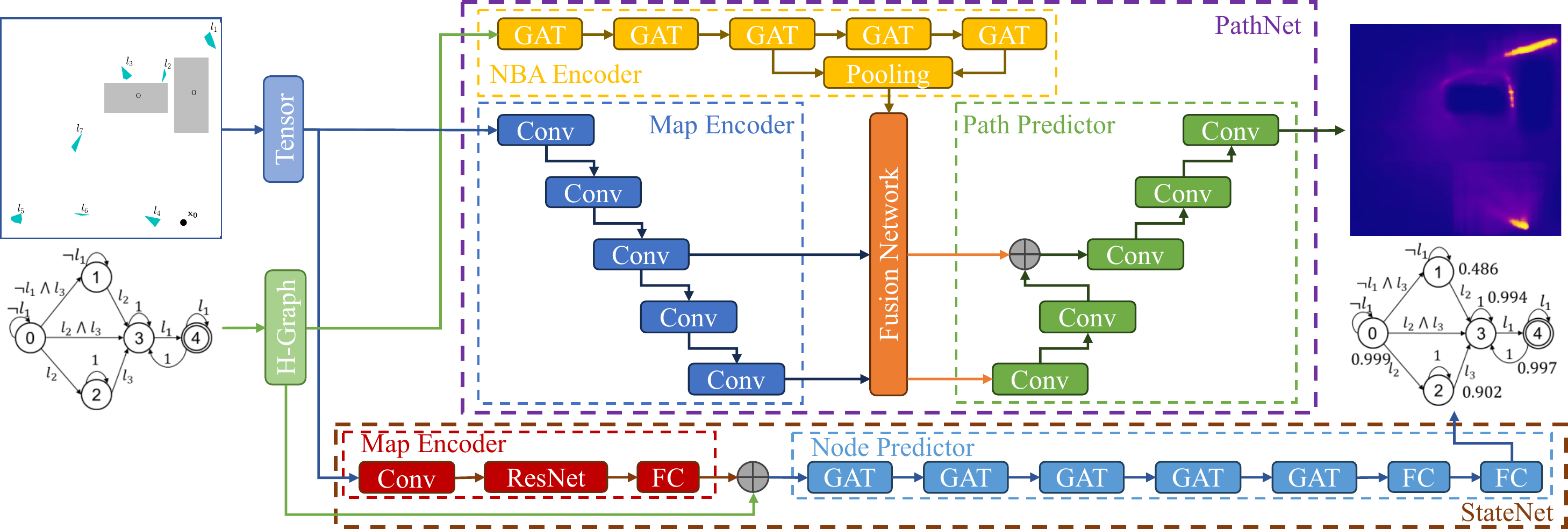}
    \caption{Overview of the sampling network.\vspace{-10pt}}
    \label{fig:model}
\end{figure*}

\subsection{Overview of  NN-Guided Approach}
\subsubsection{Purposes of Network Networks}
The main architecture of our sample net is shown in Figure~\ref{fig:model}  consisting of two sub-networks:  the Path Prediction Network ({PathNet}) and the State Prediction Network ({StateNet}). 
The inputs of the  {PathNet} and the  {StateNet} are the same, which are both the workspace map as well as the NBA for the LTL task. 
However, their purposes and outputs are different. 
Specifically, 
\begin{itemize}
    \item 
    
    The output of the {StateNet} is a vector $\mathbf{p}$ with a length of $\left | \mathcal{Q}_B \right |$. In this vector, each entry $p_i$ denotes the probability that state $i$ is involved in the optimal path. The prediction vector $\mathbf{p}$ is employed to guide the choices of $q_B^{F,\text{feas}},q_{B}^{\text{succ,1}}, q_{B}^{\text{succ,2}}\in \Q_B$ in the biased sampling.
    
    \item 
    The output of the  {PathNet} is a 
    $200\times 200$ matrix $\mathcal{P}$ 
    such that the value of each entry represents the likelihood that the entry is on the optimal path. 
    Therefore, this weight matrix $\mathcal{P}$  is used 
    as a more reasonable metric for sampling $\x^{\text{rand}}$ by considering the workspace information with preforming shortest path search. 
\end{itemize}

\subsubsection{NN-Guided Sampling Strategy}
Now, let us discuss how the proposed neural networks are used to guide the sampling process.
It still follows the basic idea of biased sampling method as detailed in steps S1-S6. 
However,  we further leverage the outputs of the two sub-networks to improve and simply the sampling process. Specifically,  at each instant, suppose that the output of the  {StateNet} and the  {PathNet} are $\mathbf{p}$ and $\mathcal{P}$. 
We set  $\alpha\in (0,1)$ as a parameter that specifies the probability of using the predicted information of  {StateNet} to guide the sampling. 
Then, to determine the sample point $\x^{\text{rand}}$, our approach makes the following changes C-1 and C-2 to steps S-1 to S-6.  
 
C-1: When selecting $q_B^{F,\mathrm{feas}}$, $q_{B}^{\mathrm {closest}}$, $q_{B}^{\text{succ,1}}$ 
    as in steps S1-S3 in the biased sampling,  
    there is a $1-\alpha$ probability that we still follow exactly the same strategy in S-1 to S-3. 
    However, there is an $\alpha$ probability that  
    all these states selected according to the probability vector $\mathbf{p}$.  In other words, we have $\alpha$ probability to activate the prediction result of the  {StateNet}.

C-2:  After obtaining state $\mathbf{x}^\mathcal{L}$ in step S-4, to sample state $\x^{\text{rand}}$,  we simplify steps S-5 and S-6 as a single step. 
    Particularly, instead of first computing the shortest path and then using the second point in the path to generate a sample distribution, here we directly use 
    the prediction result $\mathcal{P}$ by the  {PathNet} to sample $\x^{\text{rand}}$. 
    More specifically, let $(x_1,y_1)$ and $(x_2,y_2)$ be the coordinations of $\mathbf{x}^\mathcal{L}$ and $\mathbf{x}^{\mathrm {closest}}$, respectively. 
    We consider the rectangle region 
    $\textsc{Rec}=[x_1:x_2,y_1:y_2]$.  
    We define a discrete distribution over 
    all grids in $\textsc{Rec}$ according to the normalized 
    value of their weights in $\mathcal{P}$, i.e., 
    grids with larger value has more chance to be selected. 
    Then  $\x^{\text{rand}}$ is sampled randomly from the rectangle region according to this distribution.  


Here, we discuss the main features of the purposed NN-guided sampling strategy. 
First, our NN-guided approach subsumes all properties, such as probabilistic completeness and  asymptotic optimality,  of the TL-RRT* algorithm in \cite{luo2021abstraction} since we still follow the main structure of TL-RRT* and at each step, our algorithm has a non-zero probability to switch to the original sampling strategy. However, compared with the biased sampling strategy adopted in \cite{luo2021abstraction}, our NN-guided sampling strategy further jointly considers both the workspace information and the NBA, which provides a better heuristic for ``good" samples. 
Furthermore, since our strategy uses the predicted distribution directly without involving the shorstest path search for each step, its online execution is also much faster that the biased sampling strategy. 


\subsection{Inputs Encodings}\label{encoding} 
Recall that, for both the  {PathNet} and the  {StateNet}, their inputs are the workspace map and the NBA for the LTL formula.  
To leverage neural networks for processing the workspace with continuous space and NBA with graph structure, appropriate encoding techniques are needed. 

\subsubsection{Encoding for Workspace}
First, we consider the continuous workspace as a $200\times 200$ pixel image or a grid map.  
Each grid in the image corresponds to a specific point in the workspace, which is labeled, obstacle or free.  
Then  the image is  transformed into a tensor of dimensions $(m+1)\times 200 \times 200$ with $m$ be the number of different labels, i.e.,  each grid in the workspace is encoded as a vector $\mathbf{a}=[a_0,a_1,...,a_m]$. 
Specifically,  the first entry $a_0$ represents the grid's status, where  $-1$ for the initial location, $0$ for free space, and $1$ for an obstacle. 
For $i=1,\dots,m$,  we  have  $a_i=1$ if it belongs to $\mathcal{R}_i$, and  $a_i=0$ otherwise.

\subsubsection{Encoding for B\"{u}chi Automata} 
First, we convert NBA to a directed graph, where nodes and edges correspond to states and transitions, respectively, and have their features. 
Specifically, the feature for a node is a vector  $\mathbf{v}=[v_1,v_2,v_3]$, 
where 
$v_1\in \{0,1\}$ represents whether or not it is an initial state, 
$v_2\in \{0,1\}$ represents whether or not it is an feasible accepting state, and 
$v_3$ is the (normalized) distance to the closest feasible accepting state. 
For each edge, the feature is a vector $\mathbf{e}=[e_1,e_2,...,e_m]\in \{-1,0,1\}^m$ specifying the atomic propositions that need to be true for the underlying transition in the NBA.  
Since later on we   use graph neural networks (GNN) to process features of the NBA,  we  further transform the directed graph into a heterogeneous graph. 
This transformation involves adding a new node within each  original edge  so that the features of the edges are inherited by the nodes added. 
Additionally, to augment feature aggregation and spread,  a self-loop is added to each node, and a pooling node is added so that each node has a directed edge leading to this pooling node.

\subsection{Implementation Details of Neural Networks}\label{subsection_network}
%

\subsubsection{PathNet}  It has the following four building blocks.

\textbf{Map Encoder:} The purpose is to extract features from the grid map by five convolutional blocks. Each block houses a $4\times 4$ convolutional layer with a stride of $2$ and padding of $1$, complemented by a batch-norm layer, a dropout layer with probability $0.5$, and a LeakyReLU activation with a negative slope of $0.2$. Note that the sizes of the map evolves from $m \times 200 \times 200$ to $1024 \times 6 \times 6$ when passing through these blocks.  

\textbf{NBA Encoder:} 
The purpose is to extract  global features from the NBA. Comprising five layers, the encoder utilizes Graph Attention (GAT) convolutions  to address distinct edge types in the heterogeneous graph\cite{velickovic2017graph}. The node  features are proceeded by a dropout layer and a ReLU activation after convolutions. 
Finally, global mean pooling operations  are used to accumulate feature representation of the entire graph.

\textbf{Fusion Network: }
The purpose is to amalgamate features from both the workspace Map Encoder and the NBA Encoder. 
This is done by the following two steps. First, NBA features are transformed via a linear layer to attain compatibility with map features. 
Then vector concatenation are used to fuse  these harmonized features.

\textbf{Path Predictor: }
The purpose is to output the weight matrix $\mathcal{P}\in \mathbb{R}^{200\times 200}$ as the prediction for optimal paths.  
To this end, we use five up-convolution blocks to upscale and to refine the amalgamated features.
Each block is structured with a transposed convolutional layer, a batch-norm layer, a dropout layer (with probability $0.5$), and a ReLU activation. 
Drawing from the ``U-Net" architecture \cite{ronneberger2015u}, our design integrates skip connections, merging features from both the third convolutional modules of the Map and NBA Encoders. This approach capitalizes on the synergy of both encoders, enhancing the merged feature representation. Subsequent to this fusion, features are relayed to the Path Predictor via concatenation. Leveraging these connections ensures the retention of intricate spatial details alongside the depth of hierarchical features. 
Finally,  we use an $1\times1$ convolution to reshape the features to $1\times200\times200$ dimensions, and a sigmoid activation is used to refine the output.  

\subsubsection{StateNet} 
This net is essentially a classifier for the NBA states. It has a simpler structure   consisting of the following two components. 


\textbf{Map Encoder: } 
The main purpose is to encode the map into a $256$-length feature vector, which will be served as a precursor to the Node Predictor. 
Specifically, the map is processed by an $1\times 1$ convolution, transitioning the $8$-channel input to $3$ channels, thus aligning with conventional image processing frameworks.  The features then engage with the pre-trained ``ResNet-50'' model \cite{he2016deep}. 
The terminal fully connected layer of the ResNet is omitted, and is replaced by our bespoke fully connected layer, rendering the output as a $256$-length feature vector.

\textbf{Node Predictor}, tailored for node classification, starts with a GAT convolution layer, stretching the graph's node features to a 256-length dimension, in line with the map features. Each node's features are subsequently concatenated with the Map Encoder-generated map feature vector, amalgamating spatial and structural data at every node. A sequence of information dissemination follows via five sequential GAT modules, each housing a GAT convolution, a ReLU activation for non-linearity, and a dropout (with probability $0.5$) for regularization. The classification culminates with two sequential fully connected layers and a Softmax layer, analyzing the consolidated node features to yield the classification probability.

\subsection{Training Neural Networks}\label{subsection_training}

\subsubsection{Data Set Preparations}
Initially, we randomly generate $15400$ pairs of workspace and LTL formulae.  
Specifically, in each $200 \times 200$ grid map workspace, we randomly place obstacles as well as seven distinct labeled regions. The initial location of the robot is also randomized. Note that grid map are only used for the  purpose of map generation, it is still mapped to and used as  continuous workspace. 
To order to obtain an expert path for each pair of workspace and LTL task, 
we use the existing biased-sampling approach  with $10,000$ iterations. 
The obtained expert path is then encoded into a $200 \times 200$ binary matrix. 
Specifically, we mark those grids crossed by the expert path, as well as their immediate neighbors, as $1$. Then we label NBA states visited by the expert path as $1$. Finally, through data augmentations, this data set is expanded to $107800$ cases.


\subsubsection{Training Procedures}  
The training process started  with training the  {StateNet} first.
Specifically, we  first fixed  the parameters of the ResNet portion and only update the parameters of the other parts. 
When the loss became stable after 100 epochs, we unfroze the parameters of the ResNet and continued  training for another 20 epochs. 
The trainings were performed by the Adam optimizer with an initial learning rate of $0.001$ and a batch size of $128$. 
We used  cross entropy loss as the loss function. 
After training the StateNet, we trained the PathNet. 
Specifically, we  first initialized the GAT layers in the NBA Encoder using the parameters from  the GAT layers in StateNet. 
After training for 150 epochs, the loss stabilized. 
We still used the Adam optimizer with an initial learning rate of $0.0001$ and a batch size of $128$. 
We employed  the binary cross entropy loss as the loss function.

\section{Simulations \& Numerical Experiments}
In this section, we provide simulation results for the proposed method. 
First, a case study is provided to illustrate our approach. 
Then we perform a set of numerical experiments to evaluate the efficiency of our approach compared with existing sampling strategies. 
All algorithms were implemented using \texttt{Python} on a Windows 11 computer with an Intel Core i7-13700K 5.40GHz processor. \vspace{-3pt}

\subsection{Case Study}
\begin{figure}  
    \centering 
        \includegraphics[width=3.5cm]{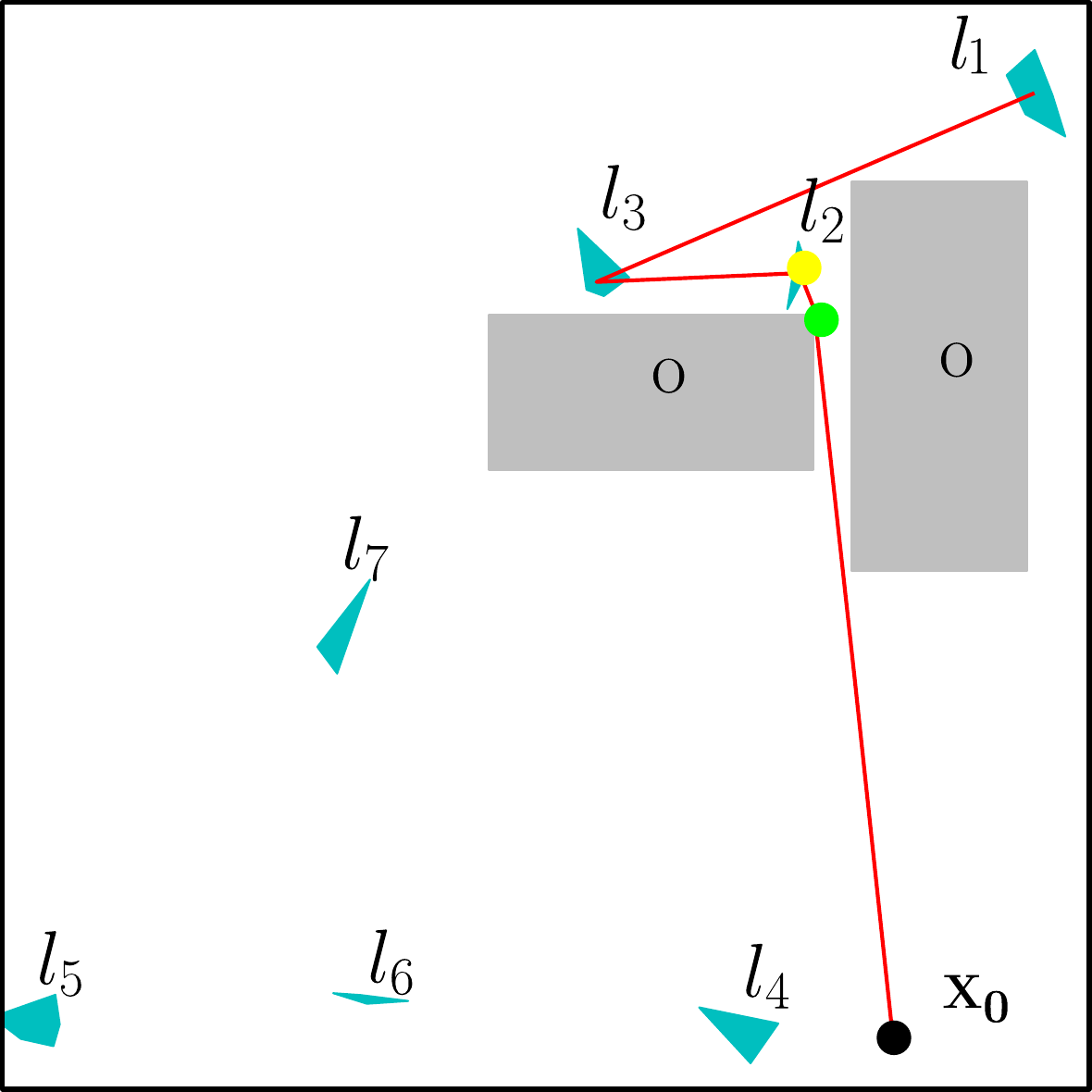}
        \caption{Workspace of the robot, where  the gray regions denote obstacles, 
        the green regions denote labeled regions, and the red lines denote the optimal path synthesized.}
        \label{fig:path}
\end{figure}

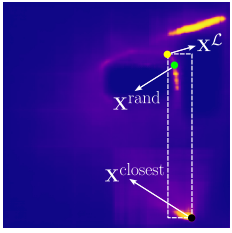
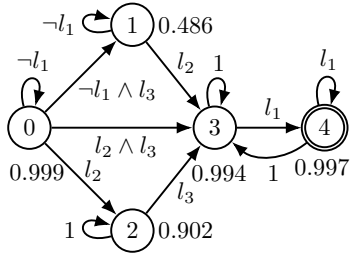
\begin{figure} 
    \begin{subfigure}{0.4\columnwidth}
    \centering  
        \includegraphics[width=3cm]{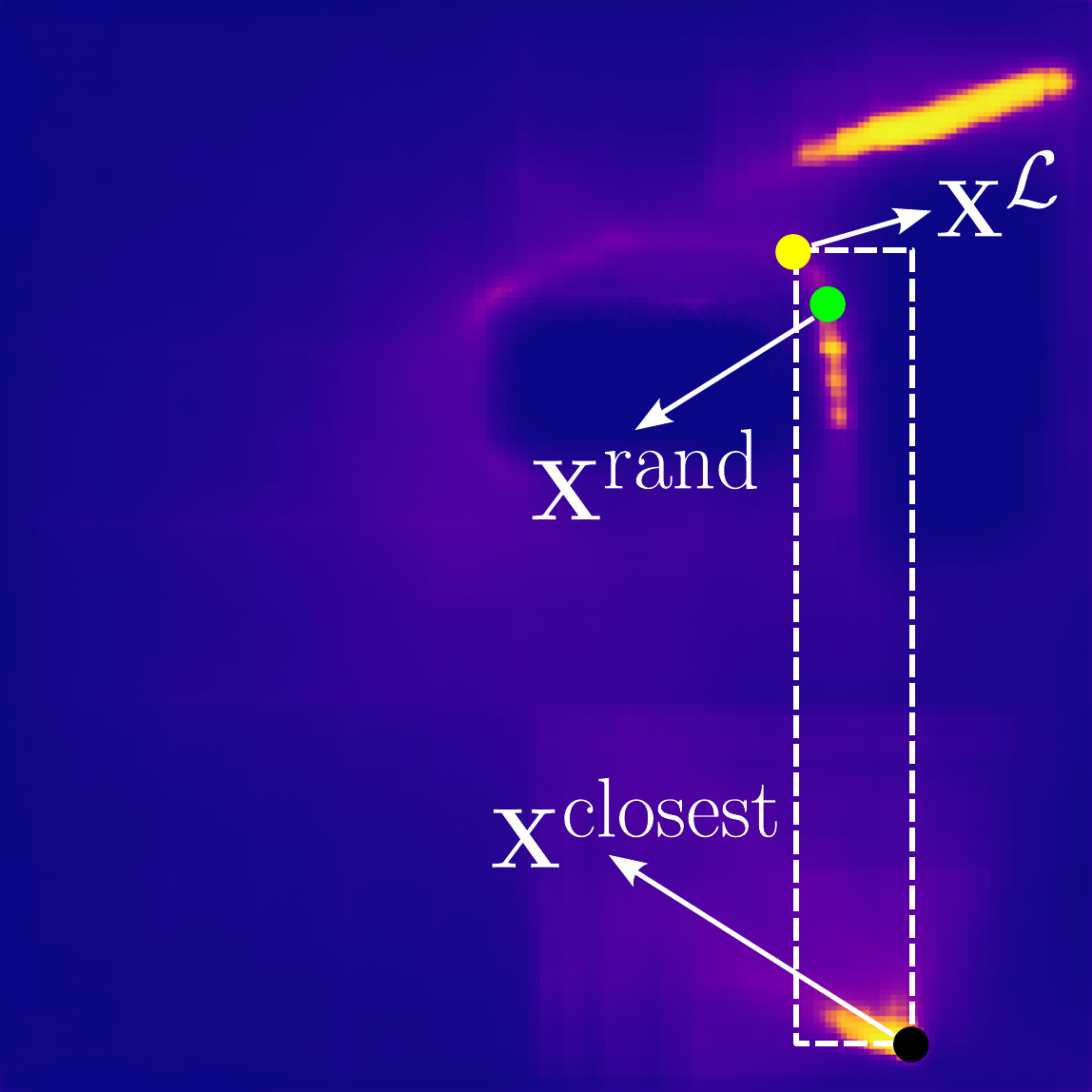}
        \caption{Output of PathNet.}
        \label{fig:prob}
    \end{subfigure} 
    \quad 
    \begin{subfigure}{0.6\columnwidth}
    \centering  
        \resizebox{1.0\linewidth}{!}{
            \begin{tikzpicture}[->,>={Latex}, thick, initial text={}, node distance=2cm, initial where=above, thick, base node/.style={circle, draw, minimum size=5mm, font=\normalsize}]		
	\node[state, base node, ] (0) at (0,0) {$0$}; 
	\node[below=of 0, xshift=0.05cm, yshift = 1.92cm] {\fontsize{10}{1} $0.999$};
	\node[state, base node, ](1) at (1.5,1.5) {$1$};
	\node[right=of 1, xshift=-2.2cm, yshift = 0cm] {\fontsize{10}{1} $0.486$};
	\node[state, base node, ](2) at (1.5,-1.5) {$2$};
	\node[right=of 2, xshift=-2.2cm, yshift = 0cm] {\fontsize{10}{1} $0.902$};
	\node[state, base node, ](3) at (2.7,0) {$3$};
	\node[below=of 3, xshift=0cm, yshift = 1.9cm] {\fontsize{10}{1} $0.994$};
	\node[state, accepting, base node, ](4) at (4.3,0) {$4$};
	\node[below=of 4, xshift=-0.1cm, yshift = 2cm] {\fontsize{10}{1} $0.997$};
	
	\path[]
	(0) edge node [right, xshift=-0.3cm, yshift=-0.2cm] {\fontsize{10}{1} $\neg l_1 \wedge l_3$} (1)
	(0) edge [loop above, looseness=8] node {\fontsize{10}{1} $\neg l_1$} (0)
	(1) edge [loop left, looseness=8] node {\fontsize{10}{1} $\neg l_1$} (1)
	(2) edge [loop left, looseness=8] node {\fontsize{10}{1} $1$} (2)
	(3) edge [loop above, looseness=8] node {\fontsize{10}{1} $1$} (3)
	(4) edge [loop above, looseness=8] node {\fontsize{10}{1} $l_1$} (4)
	(0) edge node [right, xshift=-0.2cm, yshift=0.12cm] {\fontsize{10}{1} $l_2$} (2)
	(0) edge node [below] {\fontsize{10}{1} $l_2\wedge l_3$} (3)
	(1) edge node [right, xshift=-0.2cm, yshift=0.2cm] {\fontsize{10}{1} $l_2$} (3)
	(2) edge node [right, xshift=-0.2cm, yshift=-0.2cm] {\fontsize{10}{1} $l_3$} (3)
	(3) edge node [above] {\fontsize{10}{1} $l_1$} (4)
	(4) edge [bend left=40] node [below] {\fontsize{10}{1} $1$} (3)
	;
	
\end{tikzpicture}

        }
        \caption{Output of  StateNet.}
        \label{fig:nba}
    \end{subfigure}
    \caption{Predication results of the neural networks. }
    \label{fig:case_study}
\end{figure}

We consider a robot moving in a workspace shown in Figure~\ref{fig:path} with initial position of the robot, obstacles, and labeled regions as depicted in the figure.  
We consider the following LTL task for the robot\vspace{-3pt}
\[
\phi=\square \Diamond l_1 \wedge \neg l_1 \mathcal{U} l_2 \wedge \Diamond l_3, \vspace{-3pt}
\]
i.e., the robot needs to (i) visit $l_2$ at least once without visiting $l_1$; (ii) visit $l_3$ at least once; (iii) visit $l_1$ infinitely often. 
The NBA of $\phi$ is shown in Figure~\ref{fig:nba}, where state $4$ is the unique feasible accepting state.  
The optimal plan searched by our algorithm is shown as the red lines in Figure~\ref{fig:path}. 
The robot first goes to $l_2$, then hoes to $l_3$, and finally stays at $l_1$. That is, only the prefix part contributes to the overall cost.

To better illustrate our NN-guided sampling strategy, we explain how the random tree $\mathcal{T}$ for the prefix path is expended initially from the root $q_P^0=(\mathbf{x}_0, q_B^{\mathrm{init}})$. 
Since the NBA only has one feasible accepting state, the choice for $q_B^{F,\mathrm{feas}}$ is unique, and we have $q_P^{\mathrm{closest}}=q_P^0$ since the tree only has a root so far. 
Then the StateNet predicts $\mathbf{p}=\left[0999,0.486,0.902,0.994,0.997\right]$.  
Since $0 \!\!\xrightarrow{L\left(\mathbf{x}_{0}\right)}_B\! \! 0$, we have  $q_{B}^{\text{succ,1}}=0$. From the feasible successor states of $q_{B}^{\text{succ,1}}$, we choose $q_{B}^{\text{succ},2}=2$ since it has a higher probability (0.902) than state $1$ (0.486).
Then according to the transition condition $0\xrightarrow{l_2}_B2$ in the NBA, 
we select a point in the labeled region $l_2$ as $\mathbf{x}^{\mathcal{L}}$, which is shown as the yellow point in Figure~\ref{fig:case_study}.  
Then we construct the rectangle region between $\mathbf{x}^{\mathcal{L}}$ and $\mathbf{x}^{\mathrm{closest}}$, 
and state $\mathbf{x}^{\mathrm{rand}}$ is sampled randomly according to the distribution determined by  the PathNet prediction $\mathcal{P}$  as shown in Figrue~\ref{fig:prob}.  

Note  that, in this example, there are actually two feasible prefix paths: $l_2 \to l_3 \to l_1$ and $l_3 \to l_2 \to l_1$. 
Clearly, the former is optimal with less cost.  
This information is captured by the prediction result of the StateNet, which prefers state $2\in \Q_B$ for the optimal path. 
Furthermore,  the prediction  result of the PathNet can effectively avoid obstacles and has more chance to sample near the optimal path.  
Therefore, these prediction results by the neural networks guide our search process to converge to a desired solution more quickly.\vspace{-16pt}

\subsection{Comparison with Existing Methods}

\subsubsection{Experiment Settings} 
We conduct a set of experiments to illustrate the efficiency of our NN-guided sampling strategy  compared with the existing uniform and biased sampling strategies. 
Specifically, 
independent from the training set, we generate another $240$ pairs of workspace map  and LTL task. 
For each instance, we run our method with $\alpha = 0.8$ as well as the two existing method to find a feasible plan. 
Note that, since all these RRT-based approaches are probabilistic optimal, 
we focus on comparing \emph{how fast} these strategies can find \emph{the first feasible solution}, and the performance of the first feasible solution in terms of its length. 
Formally, we consider the following metrics when   the first feasible solution is found: 
the execution time $T$ taken, 
the number of iterations $n$ required, 
the number of nodes $m$ in the random tree, and 
the length of the first feasible solution $len$.

\subsubsection{Statistic Results}
The numerical experiments results are shown in Table~\ref{table}. 
Specifically, based on the execution time $T$ of the uniform sampling approach, we divide the tasks into  simple tasks  with $T\leq \SI{180}{\second}$  and complex tasks with $T>\SI{180}{\second}$.
Then Tables \ref{table_simple} and   \ref{table_complex} show the average value of  each metric for each algorithm within these two task categories, respectively.  
Note that many complex tasks, 
the uniform sampling strategy fails to find a feasible path within \SI{2000}{\second}. For such  cases, we terminate the search by only recording  $T,n$ and $m$ without consider the length $len$.

\begin{table} 
    \centering
    \small 
    \caption{Experiment Results}
    \begin{subtable}[t]{0.45\textwidth}
        \centering
        \caption{Statistic Results for Simple Tasks\vspace{-6pt}}
        \begin{tabular}{|c|c|c|c|c|}
            \hline
            Method  & $T(\SI{}{\second})$ & $n$ & $len$ & $m$\\
            \hline
            Uniform    & 54.2475 & 1908.13  & \textbf{0.78789} & 3167.63 \\
            \hline
            Biased    & 1.03305  & 101.000 & 0.80842 & 200.307 \\
            \hline
            NN-Guided  & \textbf{0.09518} & \textbf{22.8941} & 0.90546 & \textbf{48.1395} \\
            \hline
        \end{tabular}
        \label{table_simple}
    \end{subtable}
    
     \vspace{0.1cm} 

    \begin{subtable}[t]{0.45\textwidth}
        \centering
        \caption{Statistic Results for Complex Tasks\vspace{-6pt}}
        \begin{tabular}{|c|c|c|c|c|}
            \hline
            Method  & $T(\SI{}{\second})$ & $n$ & $len$ & $m$\\
            \hline
            Uniform     & 1401.78 & 7401.27  & - & 21501.2 \\
            \hline
            Biased    & 15.5639  & 321.553 & \textbf{1.08553} & 704.082 \\
            \hline
            NN-Guided  & \textbf{2.19490} & \textbf{44.8549} & 1.12555 & \textbf{182.743} \\
            \hline
        \end{tabular}
        \label{table_complex}
    \end{subtable}\label{table}
\end{table}

The statistic results show that, for both simple and complex tasks, our NN-guided sampling strategy can  significantly enhance the  efficiency of the RRT-based algorithm.
In particular, 
the timed required to obtain a feasible solution by our NN-guided sampling strategy  
is  less than 15\% of that of the biased sampling strategy, which is already much more efficient than uniform sampling strategy. 
Furthermore, the length of the first feasible solution found by our strategy is similar to those of the other two strategies. 
We would like the remark that the metric of $len$ is not essential compared with other metrics since this is just the performance of the first feasible solution. 
The entire algorithm will converge to the optimal solution when the number of iterations increases.



\section{Conclusion}
In this paper, building upon the current state-of-the-art sampling-based LTL planning algorithms, we propose a novel sampling strategy based on multi-modal neural networks to guide the sampling process.  
Our approach, on the one hand, leverages the feature extraction power of neural networks in an end-to-end fashion, 
and on the other hand, still enjoys all good properties of the sampling-based methods such as probabilistic optimality/completeness. 
Experimental results show that our proposed sampling strategy can significantly enhance the planning efficiency of the algorithm. 
In future research, we will further improve the feature fusion methods  for multi-robot planning problems.

\newpage 

\bibliographystyle{IEEEtran}
\bibliography{references}{}

\end{document}